# Skill-Based AI Agents for Power-System Studies

Pavel Etingov and Shuchismita Biswas
PNNL
Richland, Washington

***Abstract*—This paper describes a skill-based agentic framework for power-system studies using Model Context Protocol (MCP)-connected engineering tools. A custom MCP server was developed to expose Siemens PTI PSS®E functions for power-flow analysis, dynamic simulation, result extraction, and model-validation workflows. Two implementation pathways built on a programmable OpenAI Agents software development kit (SDK) and a Claude Code command-line interface (CLI) were evaluated, both using reusable skills, subagents, MCP tools, data-repository connections, and local shell/Python execution. Both frontier-model-based implementations successfully executed representative study tasks. Success was evaluated based on task completion, output accuracy, and the need for human expert interventions. Results based on public datasets show that agentic systems can greatly accelerate power system dynamic simulation process for transmission planning studies leveraging industry-grade simulation platforms. This points toward a shift in transmission planning practice, where agentic systems could handle routine simulation setup and result extraction, allowing engineers to focus expert judgment on scenario design and interpretation rather than tool operation.**

## I. Introduction

A typical power-system planning or operation-support study requires numerous time-consuming tasks, such as preparing power-flow cases for multiple scenarios, evaluating steady-state and stability performance, configuring dynamic-simulation inputs, selecting contingencies or disturbances, assessing sensitivity cases, extracting and analyzing simulation results, and preparing engineering reports [1]. Although mature commercial and open-source tools exist for many of these steps, the end-to-end workflow remains manual, tool-specific, and dependent on expert knowledge. This creates challenges for scalability and timely decision support, and requires significant human resources and labor time. These challenges are becoming more pronounced as grid operation becomes more complex requiring engineering studies to evaluate an increasing number of operating conditions, scenarios, and uncertainty factors [2].

Recent progress in large language models (LLMs) and agentic artificial intelligence (AI) systems provides a potential mechanism for partially automating these study processes, reducing engineering time, and improving procedural consistency. Unlike earlier chat-based interfaces, in which AI assistants were primarily added as conversational front ends to existing tools, modern agent frameworks can decompose tasks, call external tools, delegate work to specialized agents, maintain context, and apply guardrails or approval checkpoints before taking actions [3]. The Model Context Protocol (MCP) has been introduced as a standardized mechanism for connecting AI agents to external tools and data sources through well-defined interfaces [4]. Recent agentic platforms also introduce reusable "skills" that package procedural instructions, metadata, scripts, templates, and tool-use guidance for domain-specific tasks [5]. Skills can significantly streamline agent development because domain knowledge can be encoded as reusable procedural modules rather than only through unstructured or case-specific prompts. Together, these developments suggest promising agentic architecture for power-system engineering, in which LLM agents do not replace validated power-system analysis and simulation tools, but instead orchestrate them through structured workflows informed by huma subject matter expertise.

Recent work has started exploring this direction for power systems [6]. A broad review of agentic AI systems for electrical power-system engineering is presented in [7], emphasizing the need for safe, reliable, and accountable agent designs for engineering applications. Several studies have demonstrated that LLMs can assist with power-system simulation tasks when supported by retrieval, tool feedback, and multi-agent decomposition. The framework in [8] demonstrated that LLMs can be adapted to perform simulation tasks using previously unseen tools, while the more comprehensive feedback-driven multi-agent framework in [9] extended this direction to broader power-system simulation studies. Reliable agentic approaches for distribution-grid analysis are further explored in [10], including adaptive retrieval, procedure generation, and supervision of tool use. Other agentic approaches have investigated the integration of LLM-based semantic reasoning with numerical solvers for power-grid violation detection and remediation [11]. Open-source efforts from the PowerAgent project, including PowerMCP and PowerSkills, have begun to expose power-system tools through MCP servers and to encode power-system procedures as reusable agent skills [12], [13]. Beyond agentic automation, related foundation-model research has explored GridFM concepts, in which models pretrained on diverse grid data and topologies may support transferable representations for power-flow-related applications and other downstream grid-analysis tasks [14].

This work is funded by the DOE Office of Electricity Transmission Reliability and Operations (TRO) Program. PNNL is operated by Battelle for the DOE under Contract DOE-AC05-76RL01830.

This paper proposes a skill-based multi-agent architecture for power-system studies using MCP-connected, industry-grade engineering tools. The implementation connects AI agents to Siemens PTI PSS®E to support representative power-system analysis tasks. In the proposed approach, an orchestration agent receives a study objective, decomposes it into subtasks, and delegates execution to specialized task agents. These task agents are supported by reusable skill files that encode power-system procedures, required inputs, tool-call sequences, validation checks, failure-handling rules, and reporting templates. MCP servers provide controlled access to PSS®E functions and power-system data.

The proposed architecture is evaluated through three representative study procedures. The first addresses power-flow and case-analysis tasks, including case loading, solution checking, data-quality review, and summary reporting. The second covers dynamic simulation, including disturbance setup, simulation execution, channel extraction, and numerical-quality checks. The third applies the play-in approach for validating power-plant and inverter-based resource models using synchrophasor or SCADA measurements [16]. Two implementation pathways are studied. The first uses the OpenAI Agents SDK, which provides Python-based control over agent orchestration, tool interfaces, and execution logic [3]. The second uses Claude Code, which enables skills, subagents, and MCP-connected tools to be configured with less custom software development [5].

## II. Agentic Implementation and Deployment Considerations

Agentic power-system use cases can be implemented using different classes of foundation models and agent runtimes. In this work, provider-hosted frontier models are used together with the OpenAI Agents SDK and Claude Code to support rapid prototyping, quick setup, and efficient comparison of agentic implementations. This choice is appropriate for an initial research study because it reduces infrastructure requirements and provides access to strong reasoning, code-generation, and tool-use capabilities. However, cloud-hosted models may raise data-security and confidentiality concerns for utility applications involving planning cases, dynamic models, sensitive measurements, or other critical-infrastructure and proprietary data. Therefore, although provider-hosted frontier models are used in this study, on-premises or locally deployed models may be preferable for future production implementations.

Open-weight or locally deployable models provide a path for secure on-premises and edge deployment, improving data control, offline operation, model-version stability, and independence from external providers. However, they require local compute infrastructure and may currently lag frontier models in reasoning performance for complex engineering tasks. The agent runtime also affects implementation. Programmable frameworks such as the OpenAI Agents SDK provide flexible control over orchestration, tools, guardrails, tracing, and MCP interfaces [3], [4], but require software-development effort. In contrast, skill-oriented environments such as Claude Code support agents, skills, and MCP-connected tools with less custom software development [5].

Other local or persistent agent concepts, including OpenClaw and NVIDIA NemoClaw-style systems, highlight the importance of self-hosted execution, sandboxing, policy controls, and privacy-preserving operation for on-premises agent deployment [17], [18]. Additional agent frameworks, such as Microsoft Agent Framework [19], LangGraph [20], and Google ADK [21], provide alternative mechanisms for multi-agent orchestration, tool use, structured outputs, and enterprise integration.

Grid-edge deployment may benefit from lightweight agents operating near local PMU or SCADA data sources. Such agents could reduce data movement, support local preprocessing and event detection, and generate compact summaries for engineering review. The model-validation use case studied here is well suited for future grid-edge deployment because disturbance data can be preprocessed locally before transfer to centralized engineering systems.

## III. Proposed Skill-Based Multi-Agent Framework

The proposed framework uses a skill-based multi-agent architecture to coordinate power-system studies through MCP-connected engineering tools. LLM agents coordinate workflows while deterministic simulation remains in validated engineering tools. Domain knowledge is encoded in reusable skills that define procedures, inputs, validation checks, tool-call sequences, error-handling instructions, and outputs. Reusable skills also reduce reliance on long, case-specific prompts, where LLMs may overlook detailed engineering instructions as the context grows. Tool access is controlled through MCP interfaces and sandboxed execution environments to support traceability and reproducibility.

### *A. MCP Server for PSS®E Tool Integration*

A central component of the proposed framework is a custom MCP server that provides controlled access to Siemens PTI PSS®E. The server is implemented in Python using the FastMCP framework, which allows Python functions to be exposed as MCP-compliant tools with structured inputs and outputs [22]. In this architecture, MCP serves as the interface layer between the LLM agents and the engineering software. Rather than allowing an agent to directly manipulate PSS®E, the file system, or operating-system commands, the MCP server exposes a defined set of approved functions. This makes the tool interface more modular, auditable, and reusable across different agent runtimes.

As illustrated in Fig. 1, the PSS®E MCP server supports the major function blocks required for power-system studies, including case management, power-flow solution, system data queries, dynamic-simulation setup, simulation execution, and result extraction. In the tested configuration, the server exposes 21 PSS®E tools. In addition to these general power-flow and dynamic-simulation functions, the server includes custom functions for model validation procedure. The validation process is based on a play-in approach, in which measured event records, such as voltage and frequency signals, are used to drive the simulation, and simulated active and reactive power responses are compared against measured responses. The model-validation functionality leverages scripts and concepts from the previously developed open-source synchrophasor analysis and power plant model validation tool suite described in [23]. MCP provides structured tool interfaces and error propagation, while skills instruct agents to report missing, corrupted, or inconsistent inputs rather than improvise unsupported values or actions.

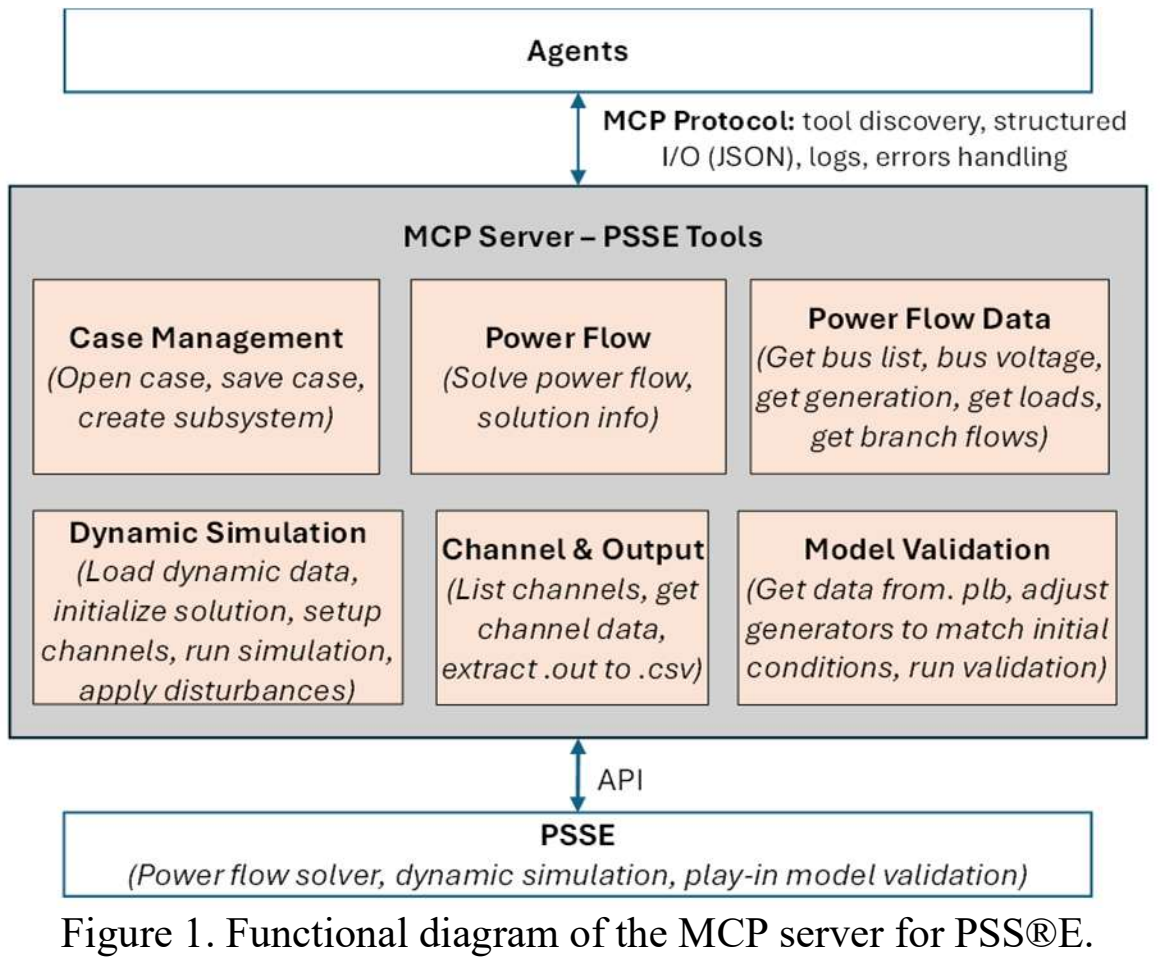

Figure 1. Functional diagram of the MCP server for PSS®E.

## B. Open AI Agents SDK Implementation

The first agentic implementation pathway was developed in Python using the OpenAI Agents SDK. The implementation includes a chat-based prompt interface through which the user submits a study request, such as solving a power-flow case, running a dynamic simulation, validating a generator model, or generating plots from simulation results. The user request is first passed to an orchestration agent, which interprets the objective, identifies the required workflow, and delegates the task to one of the specialized agents shown in Fig. 2.

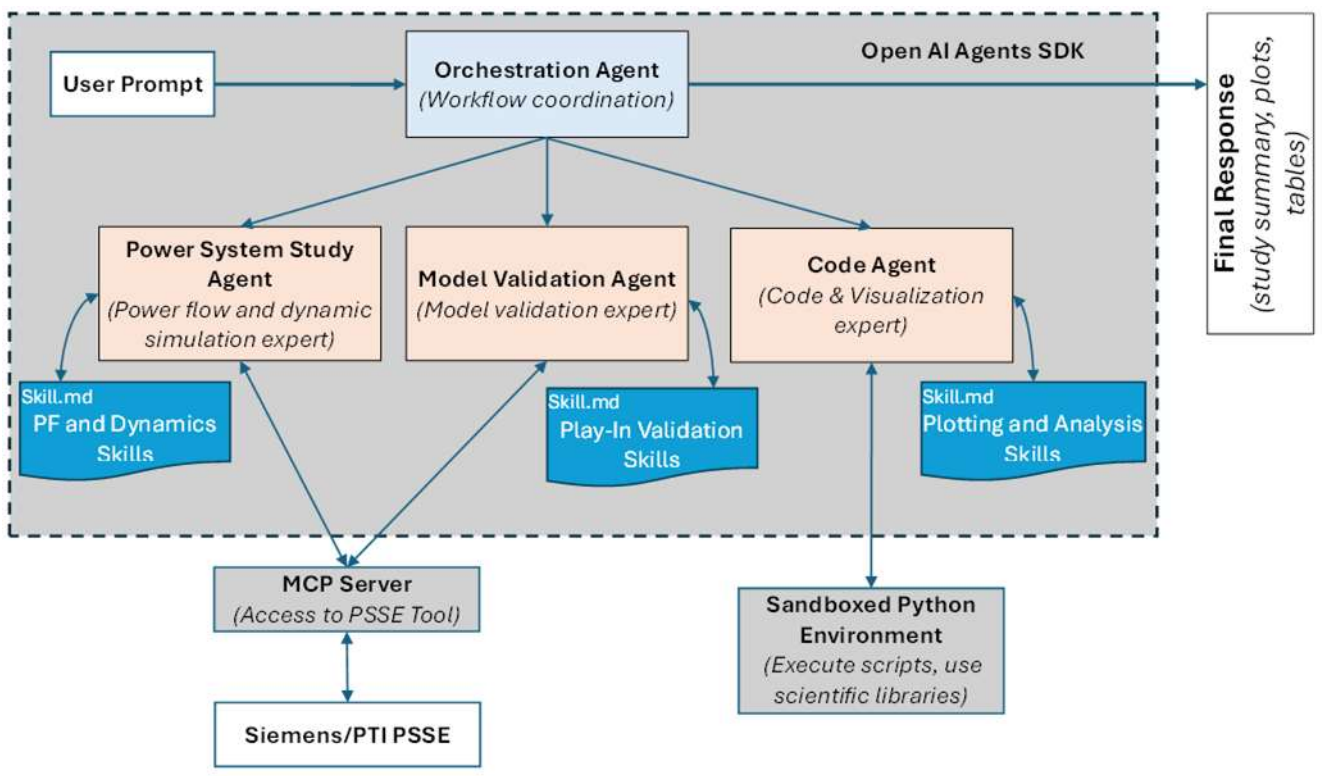

Figure 2. Multi-agent architecture for power-system studies using the OpenAI Agents SDK.

The orchestration agent coordinates the overall study process and maintains the interaction context. It determines whether the request should be handled by the power-system study agent, the model-validation agent, or the code and visualization agent. The power-system study agent performs general PSS®E-based tasks for power-flow and dynamic analysis. The model-validation agent focuses on play-in-based validation process. Both agents are associated with skills that guide the execution procedure, required inputs, tool-call sequence, validation checks, and expected outputs. A separate code and visualization agent is used for custom data analysis and plotting tasks requested by the user. The generated code is executed in a sandboxed Python environment, which separates flexible analysis and visualization from the PSS®E execution environment.

## C. Claude Code CLI Implementation

The second implementation is based on Claude Code CLI as a skill- and subagent-oriented execution environment (Fig. 3). In this implementation, Claude Code acts as the primary session coordinator. The user submits a study request through the Claude Code interface, and Claude Code determines whether to use an available skill, delegate the task to the PSS®E subagent, call tools through the MCP server, or execute local post-processing commands through its shell environment.

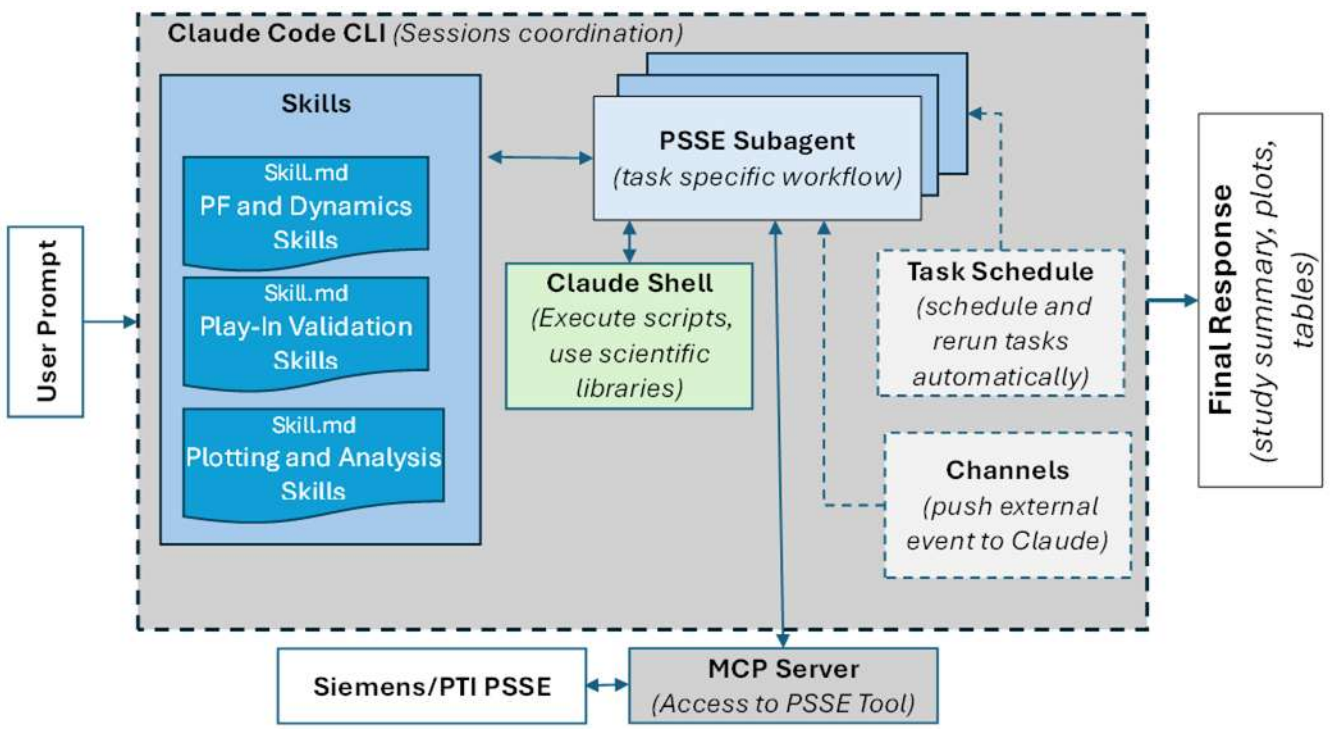

Figure 3. Agentic architecture for power-system studies using Claude Code.

The implementation uses reusable skills consistent with the OpenAI Agents SDK implementation to define procedures for power-flow and dynamic-simulation studies, play-in-based validation, and plotting or analysis tasks. A PSS®E-focused subagent is used for task-specific execution. This subagent applies the relevant study process instructions and interacts with the same PSS®E MCP server described in Section III-A. Through the MCP server, the subagent can open cases, solve power flow, load dynamics, run simulations, extract channels, and execute model-validation functions.

The Claude Code implementation requires less custom orchestration code than the Python-based OpenAI Agents SDK because session coordination, subagent invocation, skill use, shell execution, and MCP access are provided within the Claude Code environment. The same structure can also support future event-driven operation, where scheduled tasks or external channels trigger validation after detecting qualified disturbance events. We also tested a dedicated skill that connects the agent to external event data sources through application programming interfaces (APIs), including a PMU signature library [24]. This enables future workflows for retrieving and screening disturbance records and initiating domain-specific studies such as model validation or oscillation analysis.

## IV. RESULTS AND DISCUSSION

Both agentic implementations were able to execute the representative power-system study tasks considered in this paper. The OpenAI Agents SDK implementation used GPT-5.5, while the Claude Code CLI implementation used Claude Sonnet 4.6. Both showed qualitatively similar capability for the tested tasks. The evaluation is intended as an initial proof-of-concept rather than a comprehensive statistical benchmark. Although multiple repeated runs were not performed systematically, the initial results showed consistent task execution and expected output generation. Because the two implementations use different underlying foundation models, their comparison reflects practical implementation experience rather than a controlled framework or model benchmark.

A key implementation difference is that Claude Code provides an integrated agent runtime with session coordination, subagents, skills, shell execution, and MCP access, whereas the OpenAI Agents SDK requires these capabilities to be assembled and customized through Python code. Consequently, the Claude Code implementation required less custom software development and also made it easier to inspect intermediate plans, tool calls, generated scripts, logs, outputs, and execution traces. Therefore, the illustrative results below are based primarily on Claude Code outputs and screenshots.

### A. Registration and tool discovery for the PSS®E MCP server

The first step was to register the custom PSS®E MCP server with Claude Code. After registration, the server appeared in the Claude Code MCP management interface as `psse-tools` with a connected status. Claude Code was then able to discover the functions exposed by the MCP server and make them available to the agent task execution (Fig. 4).

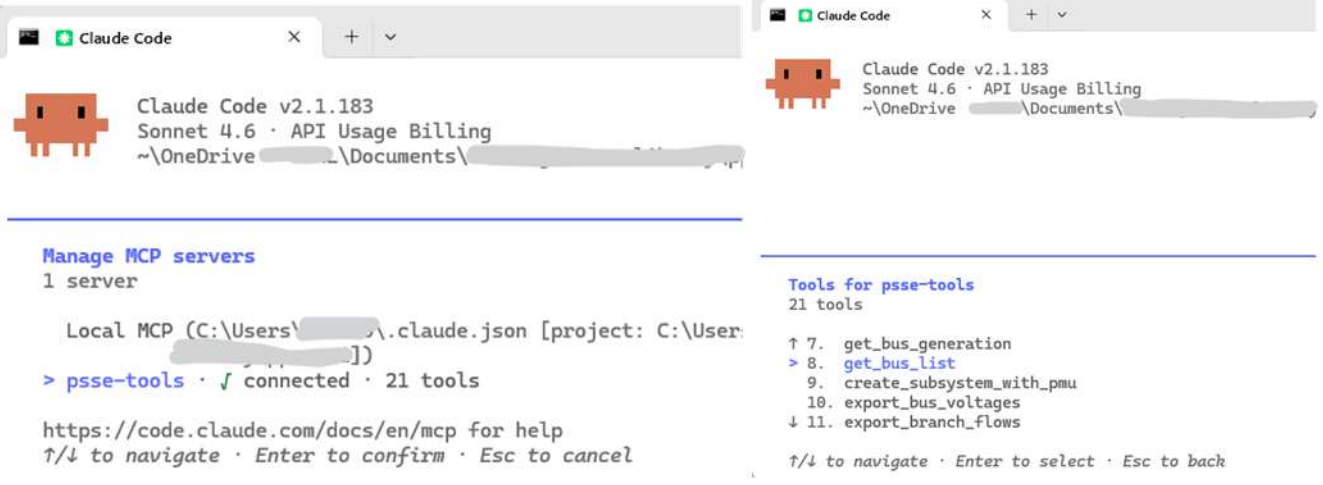

Figure 4. MCP server for PSSE setup in Claude Code.

### B. Power-Flow and Dynamic-Simulation Workflow

The system was tested on a standard transmission-planning study task that combines power-flow solution, dynamic-simulation setup, disturbance simulation, and result extraction. The user provided a single natural-language request: *"Use the Texas test-system power-flow base case and the corresponding base-case dynamic records. Set up output channels for ten generators and save voltage and frequency signals to the channel file. Run a dynamic simulation for 10 s, apply a three-phase bus fault at bus 111179 at time 5 s, clear the fault at time 5.3 s, and extract the results to CSV."*

As shown in Fig. 5, the Claude Code agent decomposed this request into a complete transient stability study. This example demonstrates that the agent can execute a multi-step engineering study from a single high-level prompt.

The next test evaluated the agent's ability to perform result post-processing and visualization from the generated simulation output. The user requested plots of the voltage and frequency signals extracted from the dynamic-simulation CSV file. As shown in Fig. 6, the Claude Code agent inspected the CSV file, identified the time column and the voltage and frequency channel columns, generated a Python plotting script, and created separate subplots for frequency and voltage. The plotting procedure also applied the frequency conversion defined in the skill file, converting frequency from per unit to hertz before generating the figure. This example demonstrates that the agent can apply domain-specific instructions encoded in reusable skills without requiring them to be repeated in the user prompt.

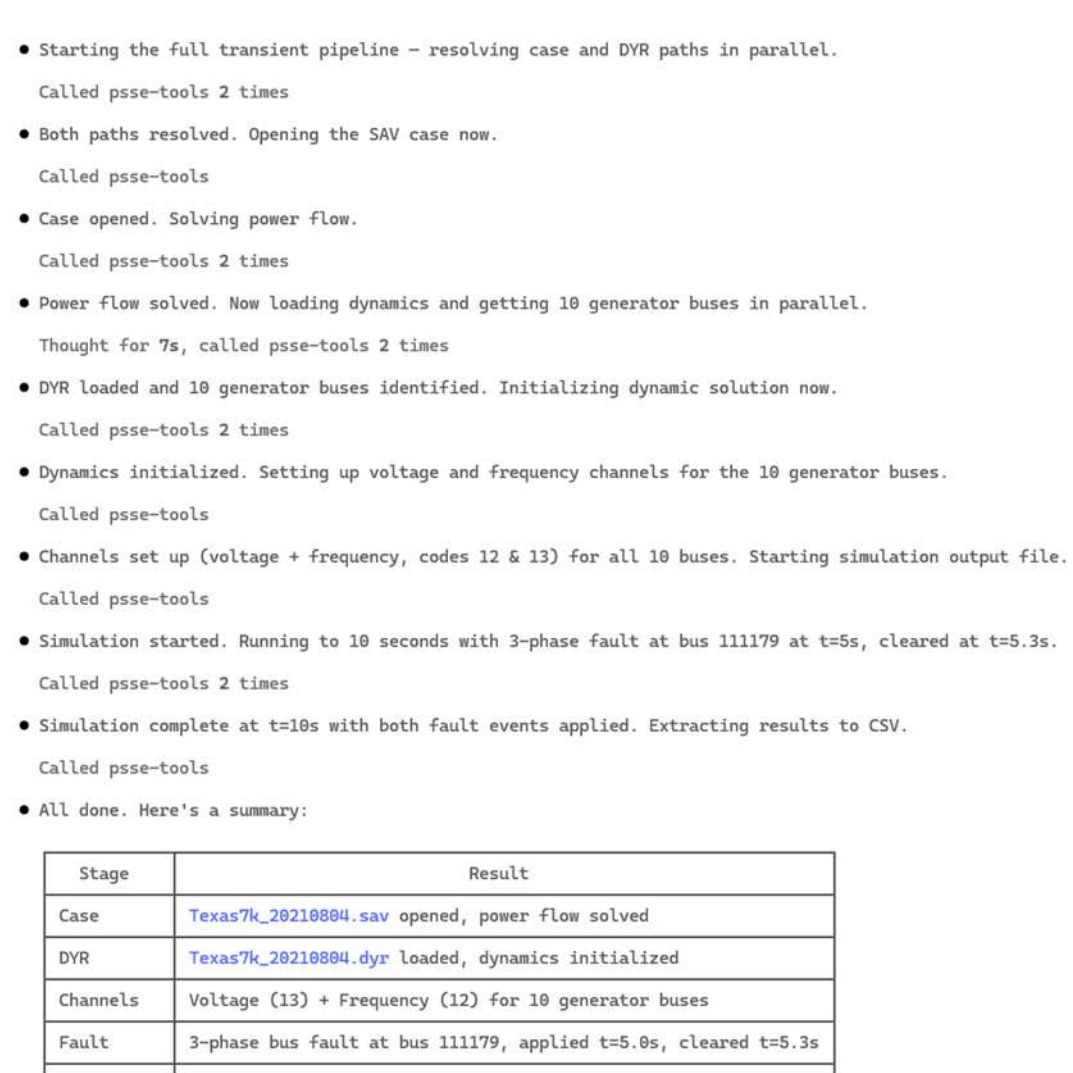

| Stage | Result |
| --- | --- |
| Case | Texas7k_20210804.sav opened, power flow solved |
| DYR | Texas7k_20210804.dyr loaded, dynamics initialized |
| Channels | Voltage (13) + Frequency (12) for 10 generator buses |
| Fault | 3-phase bus fault at bus 111179, applied t=5.0s, cleared t=5.3s |
| Simulation | Ran to t=10.0s, dt=0.01s |
| Output | agents/out/texas_fault_111179.csv |

Figure 5. Execution trace for a standard transmission-planning process.

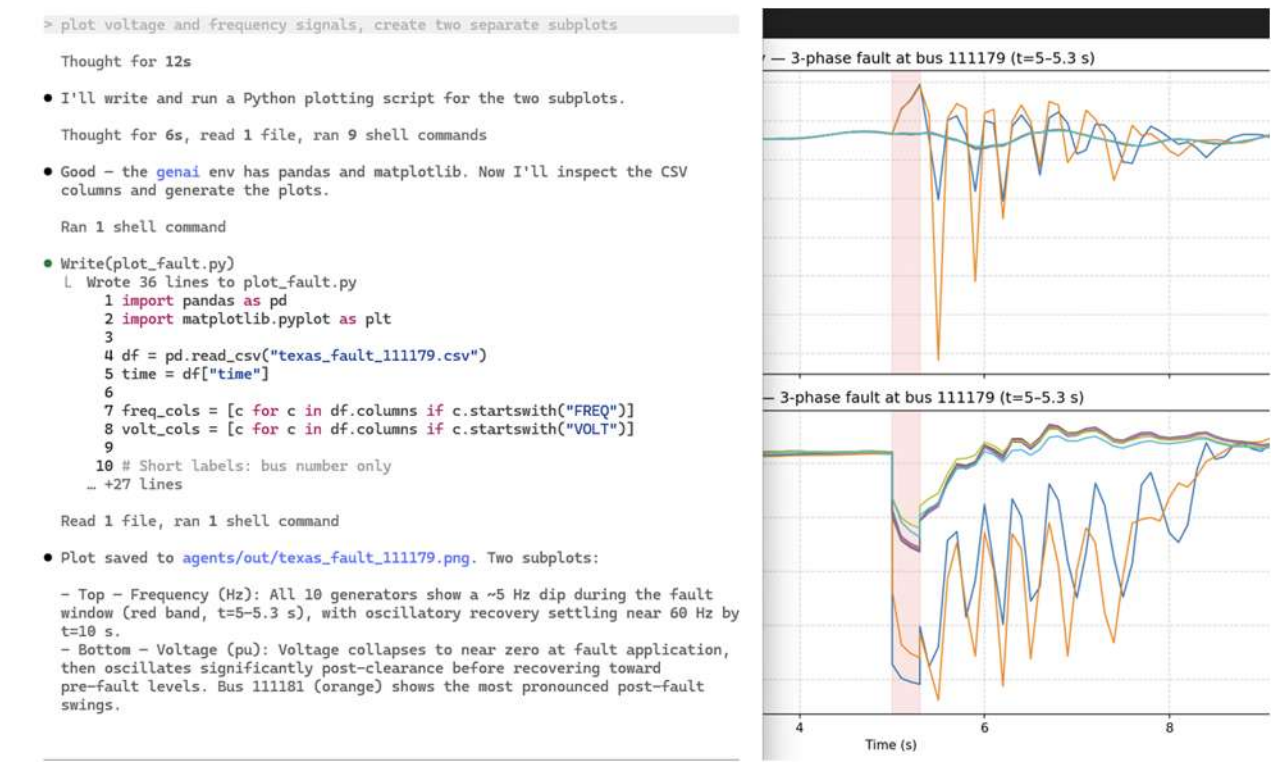

Figure 6. Agent-generated visualization of dynamic-simulation results.

### C. Power-Plant Play-In Model-Validation Workflow

A more customized use case was also tested to demonstrate the versatility of the proposed agentic framework. In this case, the agent was asked to perform a power-plant model-validation analysis based on the play-in approach [16]. Unlike the previous example, which followed a standard transient-stability procedure, this task required the agent to coordinate measured event data, dynamic-simulation outputs, calibrated and uncalibrated model responses, and comparison plotting.

As shown in Fig. 7, the Claude Code agent identified the required input files, confirmed that the dynamic-data files and play-in measurement file were available, executed the baseline validation, executed the calibrated validation, extracted the relevant output channels, and generated comparison plots. This example demonstrates that the agent can support specialized engineering workflows when domain-specific procedures are encoded in skills and exposed through MCP-connected tools.

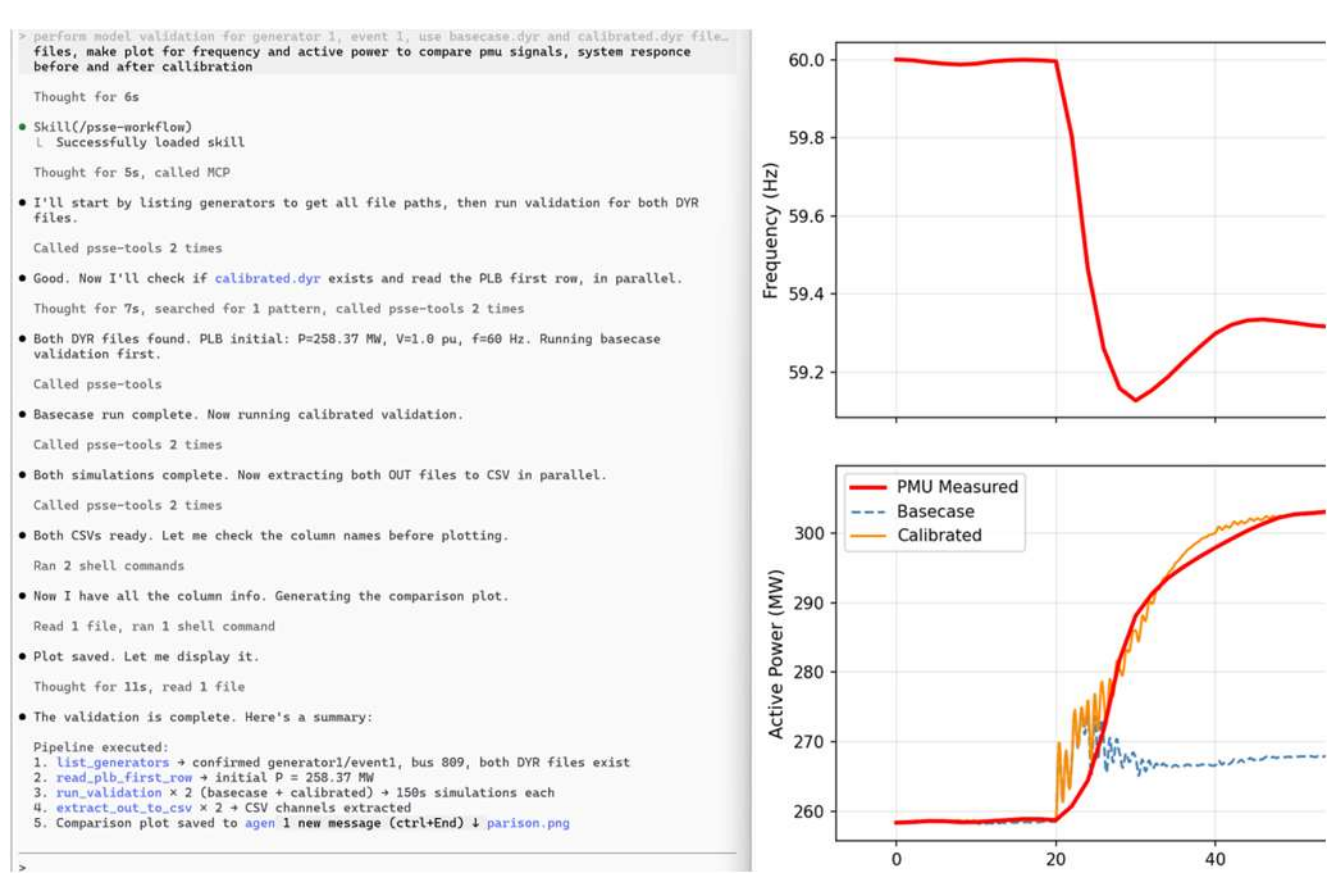

Figure 7. Execution trace and output for power plant model validation.

## V. CONCLUSIONS

This paper presented a skill-based agentic framework for power-system studies using MCP-connected engineering tools. A custom MCP server exposed selected Siemens PTI PSS®E functions for power-flow analysis, dynamic simulation, result extraction, plotting, and model-validation tasks. Two implementations were evaluated: a Python-based OpenAI Agents SDK implementation and a Claude Code CLI implementation. Both successfully executed representative study tasks, demonstrating that LLM-based agents can coordinate power-system studies through structured interfaces to deterministic engineering tools.

The results show that agentic approaches can improve the efficiency and repeatability of common power-system analysis tasks. The OpenAI Agents SDK offers greater flexibility for customized orchestration and production-style applications, while Claude Code required less custom software development and provided a simpler environment for rapid prototyping, skill use, MCP integration, shell execution, and interactive review. For the use cases studied, Claude Code was more practical when complex customization was not required.

Future work will expand MCP functions and skills to support contingency analysis, batch dynamic simulations, model-parameter calibration, oscillation analysis, and automated reporting. It will also evaluate on-premises and grid-edge deployment using locally hosted models and secure execution environments for applications involving confidential planning models and sensitive data.